\documentclass[usenatbib,usegraphicx]{mn2e}

\usepackage{times}
\usepackage{amsmath}
\usepackage{amssymb}

\newcommand{\be}{\begin{equation}}
\newcommand{\e}{\end{equation}}
\newcommand{\bear}{\begin{eqnarray}}
\newcommand{\ear}{\end{eqnarray}}

\def\aj{AJ}%
\def\araa{ARA\&A}%
\def\apj{ApJ}%
\def\aap{A\&A}%
\def\mnras{MNRAS}%
\def\prd{Phys.~Rev.~D}%
\def\zap{ZAp}%
\def\nat{Nature}%
\def\physrep{Phys.~Rep.}%
\def\Mpc{{\rm Mpc}}

\begin{document}

\title[The non-linear dark matter density field]{Exploring the
  non-linear density field in the Millennium simulations with
  tessellations -- I. The probability distribution function}

\author[Pandey et al.]{Biswajit Pandey$^{1,5}$\thanks{Email:
    biswa@mpa-garching.mpg.de}, Simon D.M. White$^1$,
  Volker~Springel$^{2,3}$ and Raul
  E. Angulo$^{4}$ \vspace*{0.2cm}\\ $^1$Max-Planck-Institut f\"{u}r
  Astrophysik, Karl-Schwarzschild-Stra\ss{}e 1, 85740 Garching bei
  M\"{u}nchen, Germany\\ $^2$Heidelberger Institut f\"{u}r
  Theoretische Studien, Schloss-Wolfsbrunnenweg 35, 69118 Heidelberg,
  Germany\\ $^3$Zentrum f\"ur Astronomie der Universit\"at Heidelberg,
  Astronomisches Recheninstitut, M\"{o}nchhofstr. 12-14, 69120
  Heidelberg, Germany\\ $^4$Kavli Institute for Particle Astrophysics
  and Cosmology, Stanford University, SLAC National Laboratory, Menlo
  Park, CA 94025, USA\\ $^5$Department of Physics, Visva-Bharati
  University, Santiniketan, Birbhum, 731235, India}

\maketitle

\date{\today}

\begin{abstract} 
  We use the Delaunay Tessellation Field Estimator (DTFE) to study the
  one-point density distribution functions of the Millennium (MS) and
  Millennium-II (MS-II) simulations. The DTFE technique is based
  directly on the particle positions, without requiring any type of
  smoothing or analysis grid, thereby providing high sensitivity to
  all non-linear structures resolved by the simulations.  In order to
  identify the detailed origin of the shape of the one-point density
  probability distribution function (PDF), we decompose the simulation
  particles according to the mass of their host FoF halos, and examine
  the contributions of different halo mass ranges to the global
  density PDF.  We model the one-point distribution of the FoF halos
  in each halo mass bin with a set of Monte Carlo realizations of
  idealized NFW dark matter halos, finding that this reproduces the
  measurements from the N-body simulations reasonably well, except for
  a small excess present in simulation results.  This excess increases
  with increasing halo mass. We show that its origin lies in
  substructure, which becomes progressively more abundant and better
  resolved in more massive dark matter halos.  We demonstrate that the
  high density tail of the one-point distribution function in less
  massive halos is severely affected by the gravitational softening
  length and the mass resolution. In particular, we find these two
  parameters to be more important for an accurate measurement of the
  density PDF than the simulated volume.  Combining our results from
  individual halo mass bins we find that the part of the one-point
  density PDF originating from collapsed halos can nevertheless be
  quite well described by a simple superposition of a set of NFW halos
  with the expected cosmological abundance over the resolved mass
  range. The transition region to the low-density unbound material is
  however not well captured by such an analytic halo model.
\end{abstract}

\begin{keywords}
  methods: data analysis - galaxies: statistics - large-scale
  structure of Universe
\end{keywords}

\section{Introduction}

One of the most important questions in cosmology is to understand the
formation of large-scale structures in the Universe. In the standard
$\Lambda$CDM model, the energy density of today's Universe is
dominated by non-baryonic cold dark matter ($\sim 23 \%$) and dark
energy ($\sim 73 \%$), whereas only $\sim 4 \%$ is comprised of all
the mass and energy associated with planets, stars, galaxies,
clusters, gas, dust and electromagnetic radiation. The study of large
scale structures is hence primarily the study of the distribution of
galaxies and the underlying dark matter.

The distribution of galaxies can be well studied with large galaxy
redshift surveys (e.g. SDSS \cite{stout}, or 2dFGRS
\cite{colles}) which provide maps of the distribution of galaxies
over large volumes in the (mostly) nearby Universe. On the other hand,
there is no direct detection of dark matter particles yet and its
existence is mostly inferred from indirect observations such as
gravitational lensing, making its observational study much harder.  In
the current paradigm of structure formation, structures form via
gravitational instability amplifying tiny density fluctuations
generated by some process in the early Universe. These initial density
fluctuations are often assumed to form a Gaussian random field. Dark
matter first aggregates hierarchically into dark matter halos and
galaxies form later in their centres by the cooling and condensation
of baryons \citep{white1978}.  As the dark matter halos form near
peaks of the initial density field, the distribution of dark matter
and galaxies on large scales is largely determined by the statistics
of these peaks \citep{bardeen1986}. On small scales, the physics of
galaxy formation complicates this picture considerably, leading to
non-linear and stochastic biasing between the distributions of dark
matter and galaxies.

Studies of the cosmic density field expected in cold dark matter
cosmologies are often based on simple and approximate analytical
models such as the halo model \citep{shethandcooray}. However,
detailed studies of the non-linear cosmic density field need to rely
on N-body simulations, which do not need to make simplifying
assumptions about the abundance and structure of halos. In the present
work, we study the non-linear density fields predicted by
high-resolution dark matter simulations, particularly the one-point
probability distribution function of the dark matter density field. We
measure this function far into the nonlinear regime and compare the
results to the halo model.

The output of N-body simulations provides the phase space distribution
of dark matter particles. Reconstructing the underlying continuous
density field represented by the discrete set of ``macro-particles''
used by the numerical scheme requires one to define an appropriate
density reconstruction scheme. For consistency, we demand that the
total mass contained in the reconstructed continuous density field has
to be exactly equal to the total mass represented by the discrete set
of particles. There exist various techniques in the literature for
density reconstruction from a given set of points which fulfill these
requirements \citep{hockney1981, silverman1986, monaghan1992,
  ascasibar2005}. The most widely used approach is to convolve the
point data with some filtering function (or simply `kernel'), yielding
a continuous map. Conventionally, the filtering function has a fixed
shape and fixed size (for example when binning particles on a regular
grid by CIC or TSC mass assignment), but this fixed smoothing
technique has the serious disadvantage that the smoothing length is
not adjusted to clustering of the particle distribution. So when a
small smoothing length is employed in order to achieve great resolving
power in high density regions like filaments and clusters, these
structure are well recovered but underdense regions like voids are
severely affected by shot noise. Conversely, if one wishes to obtain a
reasonable reconstruction of low-density voids by using a larger
smoothing length, the filaments and clusters are oversmoothed,
limiting the amount of information that can be extracted from those
regions.

A better smoothing technique is obtained by applying the SPH (smoothed
particle hydrodynamics) approach, where one employs an adaptive kernel
which adjusts itself according to the varying sampling density. For
example, the size of the (compact) kernel can be set to the distance
of the $n$-th nearest neighbor, where the value of $n$ is a
user-specified parameter. In both of these methods, the smoothing
kernel has to be specified by the user. Common choices consist of
spherically symmetric kernels, for example a simple Gaussian. However,
the fact that the geometry of the kernel is prescribed irrespective of
anisotropies present in local non-linear structures (e.g. filaments
and sheets) may introduce spurious topological signatures
characteristic of the kernel. Ideally, one would like to allow the
point distribution to decide for itself what kernel shape and size
yields the most faithful reconstruction of the local density field.
An ideal candidate for this strategy is the Voronoi tessellation
and/or its topological dual, the Delaunay tessellation
\citep{weygaert, ok, pelupessy, weygaert2007, schaap2007}. The density
estimators based on these tessellations have several advantages over
the traditional smoothing techniques which we discuss in the next
section. These advantages of DTFE over traditional smoothing
techniques for density reconstruction have made it an increasingly
popular choice in recent years (e.g. \citealt{aragon},
\citealt{zhang}, \citealt{platen}).  We will primarily use the
Delaunay tessellation field estimator (DTFE) because it offers a
parameter free reconstruction of the density field, retaining a
maximum amount of information about the density field and the topology
of structures embedded in the dark matter distribution.

The Millennium Simulation (hereafter MS) \citep{springel} is still one
of the largest high-resolution simulations of the growth of dark
matter structures. It followed the evolution of $10$ billion dark
matter particles in a $500 \, h^{-1} {\rm Mpc}$ comoving box with an
individual particle mass of $8.61 \times 10^8 \, h^{-1} \,
M_{\odot}$. The Millennium-II simulation (hereafter MS-II ;
\citealt{boylan}) simulated a $100 \, h^{-1} {\rm Mpc}$ box using the
same number of particles, thereby offering $125$ times better mass
resolution. \citet{boylan} studied the formation and statistics of
dark matter halos in the MS-II simulation. By comparing their results
with the MS simulation they found excellent convergence in the basic
dark matter halo statistics, making these two simulations ideally
suited for a study of the dark matter density field over an
unprecedented range of scales.

The one point distribution function of the cosmological density field
is one of most fundamental quantities characterizing statistical
properties of the matter distribution in the Universe. In the current
paradigm of structure formation, the present day large scale
structures grew from primordial density fluctuations with Gaussian
statistics. The one point distribution of today's density field is
however far from Gaussian as a result of gravitational evolution. In
the mildly non-linear regime, it is known that the one point
distribution of the dark matter density field obtained from N-body
simulations is reasonably well described by a log-normal distribution
\citep{coles91,kofman94,kayo,taruya}, but this approximation
eventually breaks down in the highly non-linear regime. 

In this paper, we study the one point distribution of the dark matter
density fields in the MS and MS-II using DTFE and try to interpret the
results in the simple picture provided by the halo model. The dark
matter halos are the densest sites in the cosmic mass density field,
and approximately $\sim 50 \%$ ($49.6 \%$ in the MS and $ \gtrsim 60
\%$ in the MS-II at $z=0$, with a $20$ particle limit) of the mass is
bound in resolved halos. Note that in the halo model, all of the mass
in the Universe is assumed to be part of a dark matter halo of some
mass. The density profiles of simulated CDM halos are well described
by the universal NFW profile \citep{navarro1996}, with a shape
approximately independent of mass, the amplitude of initial density
fluctuations and cosmology \citep{ navarro1997,cole,jing99}. The
concentration varies weekly with halo formation time. In the halo
model, one tries to represent the underlying dark matter distribution
as a superposition of a set of NFW halos with abundance and clustering
modelled with simplistic models or analytic fits to N-body results.

N-body simulations compute a periodic model universe of finite size
and finite mass resolution. This also requires a softening length
below which the gravitational interaction is suppressed to avoid
singularities in orbit integrations and unphysical particle
scattering. These numerical limitations are expected to influence the
ability of the simulation to resolve very high and very low density
regions, and consequently affect the tails of the one-point
distribution. The MS and MS-II use different simulation volumes, mass
resolutions and softening lengths, allowing us to study the importance
of these effects in shaping the tails of the one-point
distributions. We note that the use of a user defined kernel for
estimating densities like in SPH introduces a smoothing scale and a
corresponding resolution element which will typically have an
additional effect on the tails of the distribution. As DTFE is
self-adaptive without a free parameter, this type of effect is
expected to be less important for this scheme than other numerical
limitations due to finite volume, finite mass resolution and
gravitational softening.

This paper is structured as follows. In Section~\ref{secmethods}, we
outline the Delaunay Field Tessellation Estimator and discuss our
methodology. We then present results for toy halo models and the MS
and MS-II simulations in Section~\ref{secresults}. Finally we conclude
and summarize our findings in Section~\ref{secconclusions}.

\section{Density estimation with the DTFE} \label{secmethods}

In mathematics and computational geometry, the Delaunay tessellation
for a set of points is the uniquely defined volume-covering
tessellation with mutually disjoint tetrahedra, in which no
circumsphere of any tetrahedron contains one of the points in its
interior \citep{del, ok}. Connecting the centers of the circumscribed
spheres of neighboring Delaunay tetrahedra produces the Voronoi
tessellation of the point set, which is the topological dual of the
Delaunay tessellation. The Voronoi tessellation is a division of space
into non overlapping convex regions where each region is uniquely
assigned to one of the sampling points. All the points in these convex
regions are closer to its defining sampling point than to any other
sampling point.

\begin{figure}
\resizebox{8cm}{!}{\rotatebox{-90}{\includegraphics{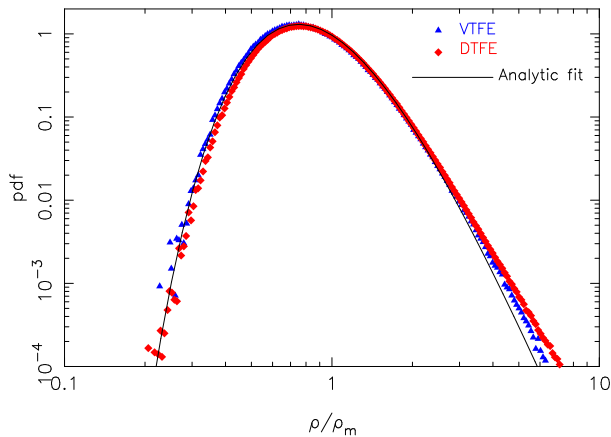}}}\\
\caption{The probability distribution functions of VTFE and DTFE
  reconstructed density fields for a Poisson point process with
  $10^{6}$ particles in 3D. The one-point distribution function in
  each bin is weighted by the bin width and the volume weighted
  density in that bin. An analytic approximation for the VTFE
  reconstructed density field given by Eqn.~(\ref{eq:analytic}) is
  shown as a solid line. The corresponding best fit values $a$, $b$,
  and $c$ for VTFE and DTFE are listed in Table~1.}
  \label{fig:1}
\end{figure}
 
\begin{table}{}
\begin{center}
\begin{tabular}{|c|c|c|}
\hline
Estimator & VTFE & DTFE  \\
\hline
$N$ &  $1000000$           &  $6772467$  \\
$a$ &  $368.56 \pm 10.75$  & $536.76 \pm 6.96$ \\
$b$ &  $7.97 \pm 0.05$     & $8.31 \pm 0.016$  \\
$c$ &  $6.03 \pm 0.031$    & $6.35 \pm 0.012$ \\
\hline
\end{tabular}
\end{center}
\caption{The number of points $N$, and the fitting parameters $a$, $b$ and
  $c$ assuming a functional form $f(\tilde{\rho})= a \,
  \tilde{\rho}^{-b} \, e^{-c/\tilde{\rho}}$ for the
  PDF of VTFE and DTFE reconstructed density fields for a 3D Poisson
  point processes. The mean is $1$ for both PDFs.  The variances of
  the PDFs of VTFE and DTFE are $0.22$ and $0.24$ respectively.}
\end{table}

Based on the geometric constructions of these tessellations, different
density reconstruction schemes can be constructed.  For example, the
density at each sampling point in the VTFE (Voronoi Tessellation Field
Estimator) is simply defined as $\rho_{i}={\, m_{i}}/{V_{i}}$, where
$m_{i}$ is the mass of the $i$-th sampling point and $V_{i}$ is the
volume of the corresponding Voronoi cell. This method assumes that the
mass of each particle is uniformly distributed inside each Voronoi
cell, keeping the density constant inside each cell. The product of
density and volume of all the Voronoi cells trivially returns exactly
the total mass of all the sampling points. But an important deficiency
of this density reconstruction is that the density field is
discontinuous at the Voronoi cell boundaries.  

An improved density estimator that addresses this deficit is the
Delaunay Tessellation Field Estimator (DTFE), which is based on a
Delaunay tessellation of the sampling points, as proposed by
\citet{schaap}. Here the density at each sampling point is defined as
$\rho_{i}={4 \, m_{i}}/{W_{i}}$ where $W_{i}$ is the volume of the
contiguous Delaunay region around the point (composed of all the
tetrahedra that have the point as one of their vertices). The sum
$W_{i}=\sum_{j} V_{ij}^{\rm Del}$ is the sum of the volumes of all
Delaunay tetrahedra that share point $i$ as one of their vertices. The
multiplication by $4$ accounts for the fact that each Delaunay
tetrahedron is contributing to the contiguous Delaunay region of four
points. The DTFE density estimation scheme assumes that the density
field inside each tetrahedron varies in a linear fashion. The gradient
of the density is assumed to be constant within each tetrahedron and
can then be computed using the density values at the four vertices of
the tetrahedron. One can then easily find the density at any other
location inside the tetrahedron using tri-linear interpolation. This
creates a continuous, volume covering, piece-wise linear density
field. It is easy to verify that the volume integral of the DTFE
density field reproduces the sum of the particle masses exactly. The
most important advantage of DTFE over conventional methods is that the
density estimates in this method do not rely on any additional
parameter. The DTFE kernel not only adapts to the local density as in
the case of SPH but also to the local geometry of the distribution.
We employ the tessellation engine of the parallel {\small AREPO} code
\citep{springel10} to construct the Delaunay mesh.


\begin{figure}
\resizebox{8cm}{!}{\rotatebox{-90}{\includegraphics{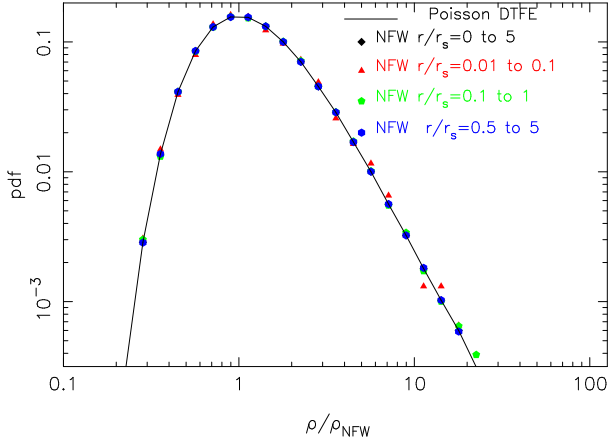}}}\\
\caption{Distributions of ${\rho}/{\rho_{\rm NFW}}$ for all the
  particles within $r_{200}$ in an NFW halo with parameters
  $M_{200}=10^{6}\,{\rm M}_\odot$, $N_{200}=10^{6}$, $r_{200}=50\,{\rm
    kpc}$ and $c=5$ (black symbols).  The distributions for particles
  residing within certain radial ranges are separately shown, as
  indicated in the figure.  Note that the distributions shown in this
  plot are simply the histograms, and their values in each bin are not
  weighted by the bin width or the volume weighted density as in
  Figure~\ref{fig:1}. The distributions in different radial bins are
  all of the same shape and are consistent with what we obtained for
  the Poisson sampling of a uniform distribution, even though the
  sampling densities vary strongly across the different radial bins.
}
\label{fig:3a}
\end{figure}

\begin{figure}
\hspace*{1cm}\resizebox{7cm}{!}{\rotatebox{-90}{\includegraphics{nfw3.ps}}}\\
\resizebox{8cm}{!}{\includegraphics{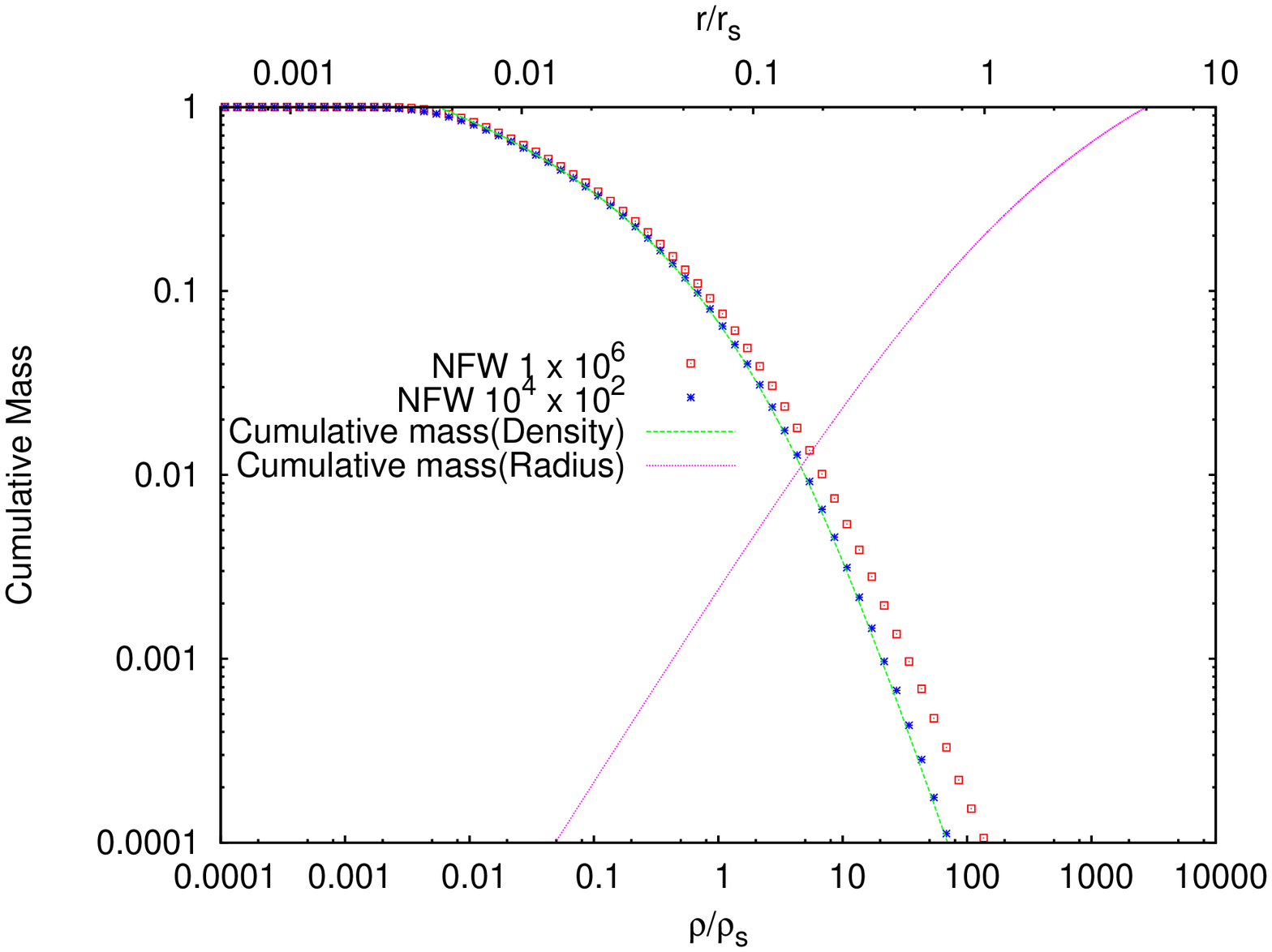}}
\caption{The top panel compares the distributions of
  ${\rho}/{\rho_{\rm NFW}(r_{s})}$ for all particles in a NFW halo
  with parameters $M_{200}=10^{6}\,{\rm M}_\odot$, $N_{200}=10^{6}$,
  $r_{200}=50\,{\rm kpc}$, $c=5$, and in all the $10^4$ halos with
  $M_{200}=10^{6}\,{\rm M}_\odot$, $N_{200}=10^{2}$, $r_{200}=50\,{\rm
    kpc}$, $c=5$. The bottom panel shows the cumulative mass as a
  function of density in the halo. The results show that the DTFE can
  quite reliably represent the mass profile of the entire halo even
  with a particle number as low as $N_{200}=10^2$. The overprediction
  of the cumulative mass at $r<r_{s}$ in the $N_{200}=10^6$ halo is
  caused by contributions from other radii due to large scatter in the
  DTFE densities. }
  \label{fig:3c}
\end{figure}

To construct and store the Delaunay mesh, {\small AREPO} uses indices
to refer to $4$ vertices and $4$ adjacent tetrahedra of each
tetrahedron which requires at least $32$ bytes of memory per
tetrahedron on $64$-bit machines (plus 4 bytes for an auxiliary
variable in practice), provided the number of points per distributed
memory region is kept low enough to allow the use of 32-bit integers
(which is the case in practice).  For a random point set there are on
average $\sim 6.77$ tetrahedra per point \citep{weygaert}, implying at
least $\sim 244$ bytes of memory per point for storing the mesh
tetrahedra. Another $20$ bytes per point are required to hold the
particle coordinates (if stored in single precision) and a unique
particle ID.  In practice, additional memory is needed for a search
tree (in order to validate individual Delaunay tetrahedra by
efficiently searching for points inside the circumsphere) and for
`ghost' cells that mesh the different tessellation patches together
across processor domains. There are more than $10$ billion particles
in both MS and in MS-II. This requires us to build a Delaunay
tessellation composed of more than $70$ billion tetrahedra for each of
these simulations. We used the {\small ODIN} machine at the Computing
Center of the Max Planck Society, Garching, to perform the mesh
constructions for MS and MS-II, using $512$ cores and $\sim 7.5$
Terabytes of memory in total. In both cases, the mesh construction
took about $20$ minutes of wall-clock time.

 \section{Results}  \label{secresults}

In cosmology, we quite often encounter Poisson sampling of an
underlying density field. For example, the galaxies in a redshift
survey or the particles in an N-body simulation can be considered as
Poisson samples for certain density ranges. Therefore it is important
to understand the impact of the Poisson sampling noise on the
statistics of the reconstructed density distribution.  Both the VTFE
and the DTFE reconstruct the density field with an adaptive spatial
resolution from a discrete set of data points. The high sensitivity of
these density estimators to the variation of local density and
geometry makes them presumably particularly sensitive to the presence
of shot noise. We therefore first examine the statistical properties
of Poisson sampling noise as seen by these density estimators.
Further, the datasets in cosmology also often involve Poisson sampling
of highly inhomogeneous distributions for example from N-body
simulations and galaxy surveys. It is also important to test whether
the one point distribution for uniform Poisson sampling of a
homogeneous distribution also describes the noise caused by Poisson
sampling an inhomogeneous distribution. We address this question with
Monte Carlo simulations of NFW halos.

We start this section by characterising the performance of the DTFE
for a Poisson sample (in Subsection~\ref{subsec1}) and for Monte Carlo
realisations of NFW halos (in Subsection~\ref{subsec2}). These test
cases, for which the underlying continuous density field is known,
will help us to understand and interpret the main results of the
paper, the one point density distributions of the cosmic density field
given by the DTFE applied on the Millennium simulations (in
Subsection~\ref{subsec3}). We finalise this section by examining the
impact in our results of halo substructure and ellipticity, as well as
of numerical setup of the simulations.

\subsection{Density PDF of a 3D Poisson point process} \label{subsec1}

We begin by studying the one-point distribution function of the
density for a Poisson process analyzed with Voronoi tessellations. A
Poisson Voronoi tessellation results when the generating points of the
Voronoi cells are a Poisson point sampling of a uniform field. In the
case of one-dimensional Voronoi tessellations, one can rigorously
derive the probability distribution of the lengths of the segments,
which is given by
\begin{equation}
g(\tilde{x})=4\, \tilde{x} \, \exp(-2 \, \tilde{x}),
\end{equation}
where $\tilde{x}={l}/{\left<l\right>}$. Here $l$ is the length of
the Voronoi cell and $\left<l\right>$ its average. No analytical
results are known for the size distributions of Poisson Voronoi cells
in 2D and 3D. Empirical studies using Monte Carlo realizations fit the
distribution of surface area or volume of the Voronoi cells in 2D or
3D with a gamma type probability distribution function \citep{kiang}
approximated by
\begin{equation}
g(\tilde{x};a)=\frac{a^a}{\Gamma(a)}\tilde{x}^{a-1}e^{- a \,
  \tilde{x}},
\label{eq:approx}
\end{equation}
where $\tilde{x}={v}/{\left<v\right>}$ is the size of the Voronoi cell
in units of the average cell size, $0 \ge \tilde{x} \ge \infty$, and
$a$ is a constant whose value depends on the dimensionality of the
space. Monte Carlo experiments suggest $a=2$, $4$ and $6$, for 1D, 2D
and 3D, respectively.

The probability of a random point to lie inside a Voronoi cell of size
$\tilde{x}$ is the product of $g(\tilde{x})$ and $\tilde{x}$, which in
other words is the probability of a random point to have a density
$\tilde{\rho}={1}/{\tilde{x}}$ in the VTFE reconstructed density
field. Following this definition, the one point distribution function
of the VTFE reconstructed density field in 3D is
\begin{equation}
{\rm d}g(\tilde{\rho})=dg(\tilde{x}) \times \tilde{x}=388.8 \, \tilde{\rho}^{-8} \, e^{-6/\tilde{\rho}} \,
{\rm d}\tilde{\rho} ,
\label{eq:analytic}
\end{equation}
if the hypothesis that $g(\tilde{x};a)$ follows equation
(\ref{eq:approx}) with $a=6$ is indeed correct. To verify these
results, we generate a Poisson point process with $10^6$ points in a
cubic box $100 \, h^{-1} \, \Mpc$ on a side. We construct the Voronoi
tessellation and Delaunay tessellation of the points and then estimate
the density field values at the sampling points using both VTFE and
DTFE. We compute the one point distribution of both the VTFE and DTFE
reconstructed density fields. We use only the density estimates at the
sampling points for computing the one-point distribution of the VTFE
reconstructed density field, and determine the best-fit parameters
assuming a functional form $f(\tilde{\rho})= a \, \tilde{\rho}^{-b} \,
e^{-c/\tilde{\rho}}$ to describe the one-point PDF of the VTFE
reconstructed density field.  In the VTFE density reconstruction
scheme, the density at the location of the sampling points is defined
as the inverse of the volume of its corresponding Voronoi cell
weighted by its mass, whereas in the DTFE scheme the volumes of the
contiguous Delaunay cells are used instead.

In the DTFE density reconstruction, the density field inside each
tetrahedron varies linearly, whereas in the VTFE density
reconstruction the density inside each Voronoi cell is constant. To
account for this difference and arrive at a more appropriate
comparison between DTFE and VTFE, we randomly select one point inside
each Delaunay tetrahedron through uniform sampling and determine its
density estimate by linearly interpolating from the four vertices of
the corresponding tetrahedron. We weight each such point by the volume
of the corresponding tetrahedron. We also weight each point in VTFE by
the volume of their Voronoi cells. As there are on average $\sim 6.77$
Delaunay tetrahedra per point for a random data set \citep{weygaert},
we get $6772467$ density estimates from the DTFE reconstructed density
field by randomly choosing one point from each tetrahedron. The
one-point distributions of the VTFE and DTFE reconstructed density
fields are both fitted to the same functional form of equation
(\ref{eq:analytic}) and the best-fit parameters are listed in
Table~1. We find that the distributions obtained for the VTFE density
field are well described by equation (\ref{eq:analytic}), and the
values of our best-fit parameters approach the literature values as
the number of particles is increased. The distribution obtained for
the DTFE density field is pretty similar overall, but it is clearly
not exactly the same. Both for the VTFE and DTFE reconstructions,
equation (\ref{eq:analytic}) noticeably underpredicts the high density
tail of the one-point distribution function. It is also interesting
that the two schemes have very similar variances despite the seemingly
larger smoothing involved in the DTFE scheme.

\begin{figure*}
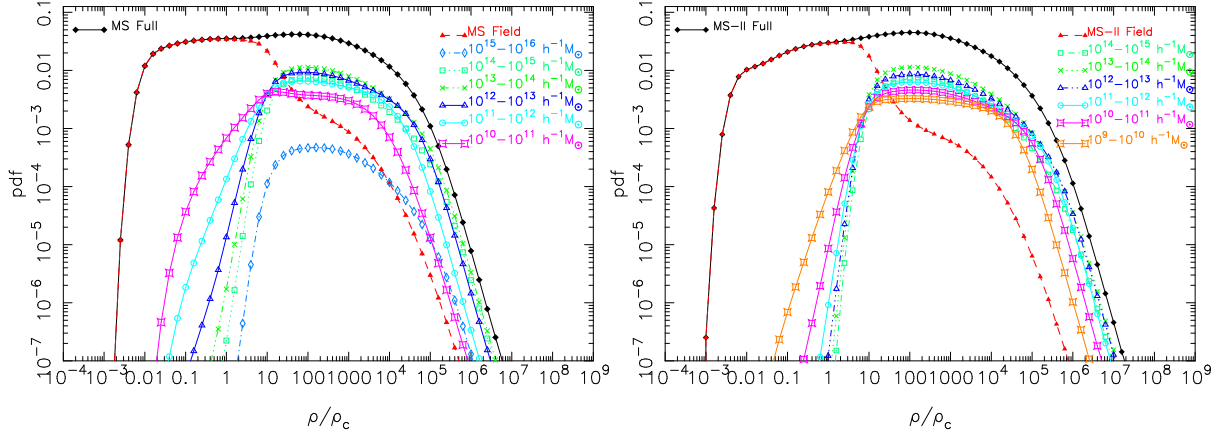

\resizebox{8cm}{!}{\rotatebox{-90}{\includegraphics{mill_all.ps}}}%
\resizebox{8cm}{!}{\rotatebox{-90}{\includegraphics{mill2_all.ps}}}
\caption{The left and right panels show the density distributions
  constructed with DTFE for the MS and the MS-II, respectively. After
  computing the DTFE density for all the particles, we identify the
  particles residing in FoF halos of different mass ranges, and the
  particles which do not reside in any halo, and then compute the
  density distribution for each of the components separately. The
  distributions for non member particles and for particles in
  different halo mass bins are shown with different colours and
  symbols (as indicated in the panels) together with the distribution
  for all the particles (black squares). }
  \label{fig:4}
\end{figure*}

\begin{figure*}
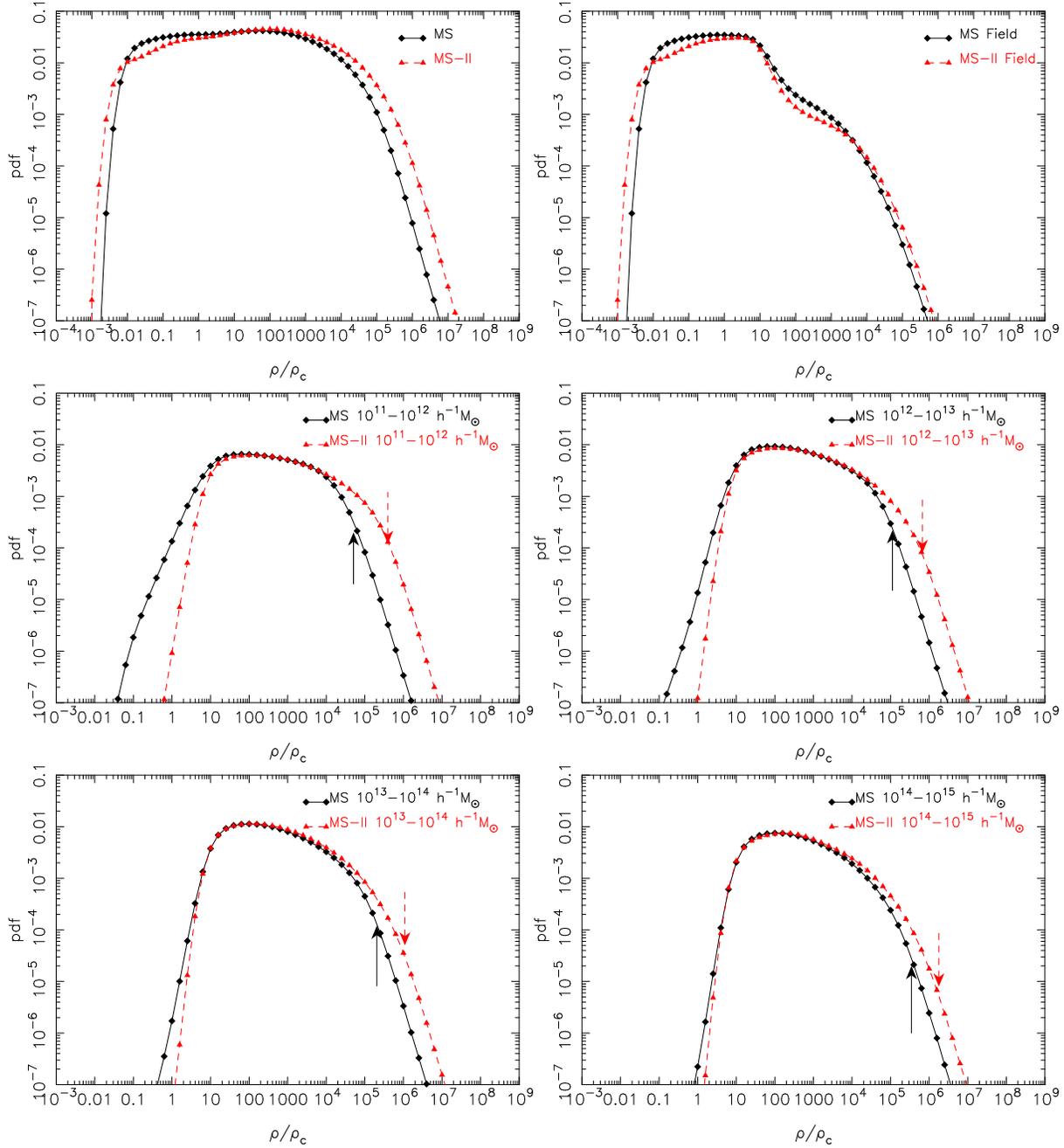

\resizebox{8cm}{!}{\rotatebox{-90}{\includegraphics{mill_mill2_full.ps}}}%
\resizebox{8cm}{!}{\rotatebox{-90}{\includegraphics{mill_mill2_field.ps}}}\\
\resizebox{8cm}{!}{\rotatebox{-90}{\includegraphics{mill_mill2_11t12.ps}}}%
\resizebox{8cm}{!}{\rotatebox{-90}{\includegraphics{mill_mill2_12t13.ps}}}\\
\resizebox{8cm}{!}{\rotatebox{-90}{\includegraphics{mill_mill2_13t14.ps}}}%
\resizebox{8cm}{!}{\rotatebox{-90}{\includegraphics{mill_mill2_14t15.ps}}}\\
\caption{Comparison of DTFE density distributions for all the
  particles, all the non-halo particles, and all the particles in
  different halo mass bins in MS and MS II. The bottom four panels
  show different halo mass ranges, as labeled. The arrows in each
  panel show the density corresponding to the softening radius in MS
  and MS-II for each mass bin. The top left shows all the components
  together (i.e.~all the particles in the simulations) while the top
  right compares only the non-halo particles.}
  \label{fig:7}
\end{figure*}

\begin{figure*}
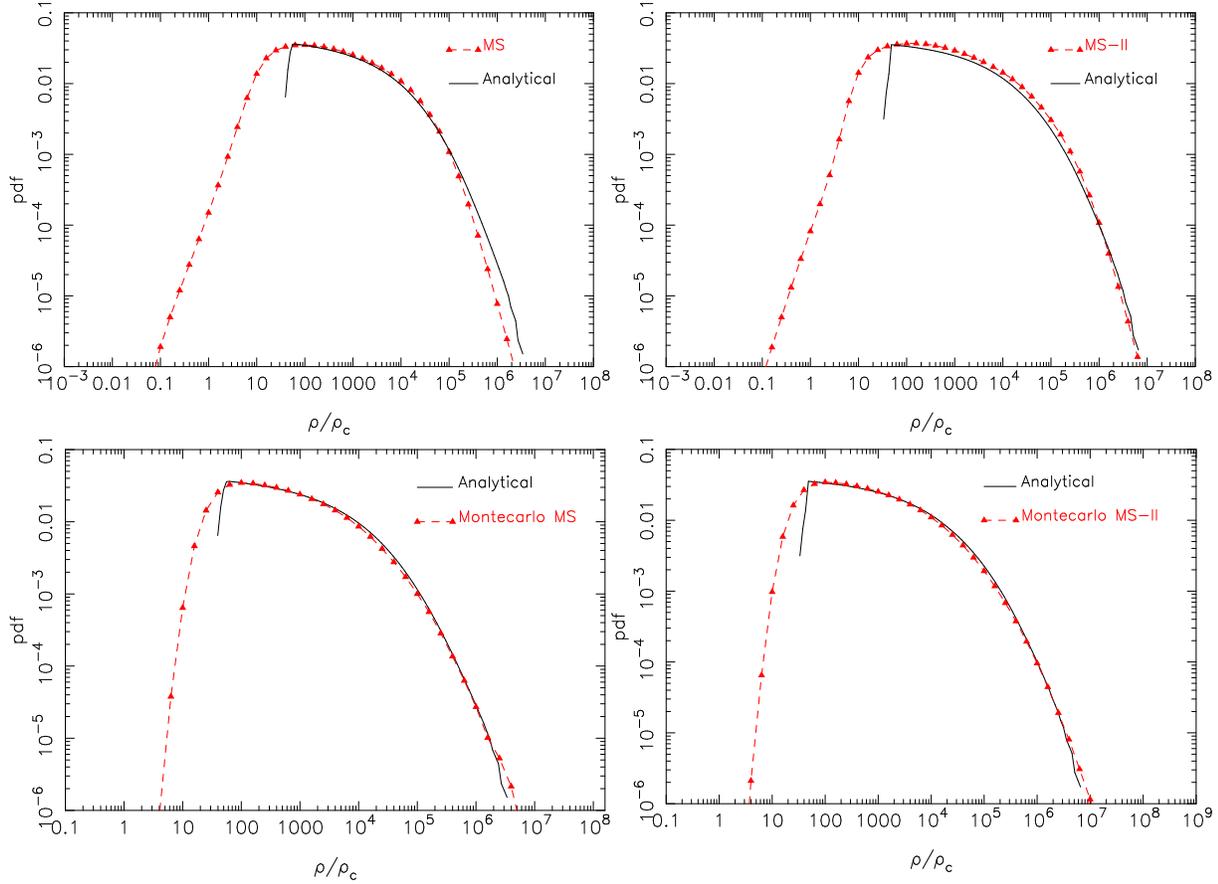

\resizebox{8cm}{!}{\rotatebox{-90}{\includegraphics{nbcompare1.ps}}}%
\resizebox{8cm}{!}{\rotatebox{-90}{\includegraphics{nbcompare2.ps}}}\\
\resizebox{8cm}{!}{\rotatebox{-90}{\includegraphics{analytic_compare_mill.ps}}}%
\resizebox{8cm}{!}{\rotatebox{-90}{\includegraphics{analytic_compare_mill2.ps}}}\\
\caption{Comparison of our analytical halo model predictions for the
  one-point distribution of the density with the direct numerical
  simulation results summed over an equal range of halo mass bins (top
  row), or with the similarly summed Monte Carlo realizations of NFW
  halos over the same mass range (bottom row). For the MS we carry out
  the sum over $5$ equally spaced different halo mass bins, each
  spanning a decade in the halo mass range $10^{11}- 10^{16}\,h^{-1}
  M_{\sun}$.  Similarly, for MS-II the sum is carried out over $6$
  equally spaced different halo mass bins in the halo mass range
  $10^{9}- 10^{15}\,h^{-1} M_{\sun}$. In both cases, we show results
  for the MS (left panels) and the MS-II (right panels) as red dashed
  lines, while the analytical halo model is shown with solid black
  lines.}
  \label{fig:8}
\end{figure*}

\begin{figure*}
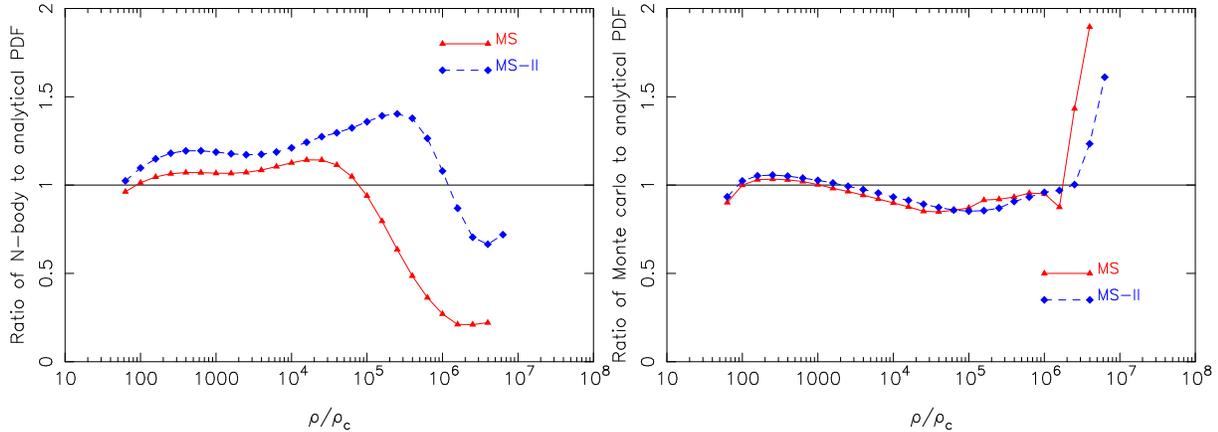

\resizebox{8cm}{!}{\rotatebox{-90}{\includegraphics{ratiopdf1.ps}}}%
\resizebox{8cm}{!}{\rotatebox{-90}{\includegraphics{ratiopdf2.ps}}}\\

\caption{Same as Fig.~\ref{fig:8}, but here we show the ratios of the
  one-point distribution obtained from N-body simulations to our
  analytical halo model predictions (left panel), and similarly for
  Monte Carlo realizations of NFW halos relative to our analytical halo model
  predictions (right panel).}
  \label{fig:8a}
\end{figure*}

\subsection{Monte Carlo realizations of NFW halos} \label{subsec2}

Dark Matter halos in N-body simulation are highly inhomogeneous
systems which we idealize as systems with spherical NFW density
profile \citep{navarro1996, navarro1997}.

The NFW density profile is given by
\begin{equation}
\rho(r)=\frac{\rho_{c} \, \delta_{c}}{\frac{r}{r_{s}} \, \left(1+\frac{r}{r_{s}}\right)^2},
\label{eq:nfw}
\end{equation}
corresponding to a cumulative mass within radius $r$ of
\begin{equation}
M(r) =\int_{0}^{r} 4  \pi  r'^2 \rho(r')\,{\rm d}r'
 = 4  \pi  \rho_{c}  \delta_{c}  r_{s}^3  \left[\ln(1+\tfrac{r}{r_{s}})-\tfrac{{r}/{r_{s}}}{1\,+\,{r}/{r_{s}}}\right].\\
\label{eq:cumulmass}
\end{equation}
Here $\rho_{c}={3 H_{0}^{2}}/({8 \pi G})$ is the critical density, 
$r_{s}= {r_{200}}/{c}$ is the scale radius, $r_{200}$ is the
virial radius, $c$ is the concentration of the halo, and $\delta_{c}$
is the characteristic density. The virial mass of the halo is
$M_{200}=200 \, \rho_c \, (4\pi/3) \, r_{200}^{3}$. The characteristic
density
\begin{equation}
\delta_{c}=\frac{200}{3} \,
\frac{c^3}{\ln(1+c)-{c}/{(1+c)}}
\end{equation}
is related to the concentration $c$ by the requirement that the mean
density within $r_{200}$ should be $200$ times the critical density.

To generate our mock NFW halos, we first specify $M_{200}$ and the
particle mass. This determines the number $N_{200}$ of particles which
reside within $r_{200}$. Next, we specify the concentration of the
halo using the mass-concentration relation determined from the
Millennium Simulation by \citet{neto}.  We then populate the halo with
particles using a Monte Carlo sampling technique, i.e.~the probability
to place a particle at a certain radius is made proportional to ${\rm
  d}M$ from equation (\ref{eq:cumulmass}). This is augmented with
isotropically selected angular co-ordinates $\theta$ and $\phi$.  To
avoid boundary effects, we extend the NFW halo radially out to $3 \,
r_{200}$.

\subsubsection{One-point distribution in different parts of the same halo}

We generate Monte Carlo realizations of an NFW halo with the following
parameters: $M_{200}=10^{6} {\rm M}_\odot$, $N_{200}=10^{6}$,
$r_{200}=50\, {\rm kpc}$ and $c=5$. The halo extends out to $3 \times
r_{200}$. We put the halo at the center of a cubic box with a side of
length $6 \times r_{200}$ and construct the Delaunay tessellation
using all the particles in the halo. Note that this leaves empty
regions at the corners of the box. The Delaunay cells near the halo
boundary will then be very extended due to the presence of these empty
regions resulting in spurious density estimates. In order to avoid
these boundary effects we limit our analysis to the particles residing
within the virial radius $r_{200}$ of the halo. We choose three radial
bins, ${r}/{r_{s}}=0.01-0.1$, ${r}/{r_{s}}=0.1-1$ and
${r}/{r_{s}}=0.5-5$, in order to probe different regimes of sampling
density. We identify all the particles residing in these radial ranges
and compute the ratio of the DTFE estimate of density to the NFW
expected value for each particle in each radial bin.  The results for
the three different radial bins together with that for all the
particles within $r_{200}$ are shown in Figure \ref{fig:3a}. Despite
the fact that the different radial bins have different sampling
density, the one-point distributions in different radial bins are all
the same and are consistent with that obtained for a Poisson point
sampling of a uniform distribution.


\subsubsection{Dependence on mass resolution}

Halos identified in N-body simulations (or galaxy groups found
in surveys) consist of different numbers of particles.  In order to
understand how the one-point distribution of DTFE reconstructed
density fields depends on the number of particles used to resolve
them, we generate a set of five NFW halos each with the same parameters
$M_{200}=10^{6} {\rm M}_\odot$, $r_{200}=50 {\rm kpc}$ and $c=5$, but
with different numbers of particles: $N_{200}=10^{6}$, $10^{5}$,
$10^{4}$, $10^{3}$, and $10^{2}$, respectively. The one-point
distribution function of the DTFE reconstructed density field of a
$N_{200}=10^2$ halo will be noisier compared to a $N_{200}=10^6$ halo
due to effects of discreteness. In order to take the discreteness
effects into account we generate different numbers of NFW halos for
each resolution: $1$ with $N_{200}=10^{6}$, $10$ with
$N_{200}=10^{5}$, $10^{2}$ with $N_{200}=10^{4}$, $10^{3}$ with
$N_{200}=10^{3}$ and $10^{4}$ with $N_{200}=10^{2}$ particles. One
then has same total number of density estimates for each resolution,
allowing a straightforward comparison of the one-point distribution
functions at the same noise level.

In the top panel of Figure~\ref{fig:3c}, we show the one-point
distribution of ${\rho}/{\rho_{\rm NFW}(r_{s})}$ for all the particles
within $r_{200}$ for NFW halos with $N_{200}=10^{6}$ and
$N_{200}=10^{2}$ particles. The distributions look very similar, except
that the high density tail of the distribution for the
$N_{200}=10^{2}$ halo shows a slight shift towards lower density as
compared to the $N_{200}=10^{6}$ halo.

We also compute the cumulative mass fractions as a function of density
from the distributions of the $N_{200}=10^{2}$ and $N_{200}=10^{6}$
halos, and compare them in the bottom panel of
Figure~\ref{fig:3c}. The theoretical prediction is shown with a solid
line.  This quantity gives the fraction of the virial mass contained
when the NFW density profile is integrated from the center of the halo
up to a certain density. The cumulative mass fraction as a function of
radius is also shown, by indicating the radius along the $x$-axis on
top.  Interestingly, the plot shows that even with $N_{200}=100$ one
can quite reliably reproduce the cumulative mass fraction from the
measured one-point density distribution of the NFW halo. The
$N_{200}=10^{6}$ halo overpredicts the cumulative mass at
$\rho>\rho_{s}$ which roughly corresponds to $r<{r_{s}}/{2}$.  This
overprediction is related to the large scatter in the DTFE density
estimates which causes a broad range of radii in the halo to
contribute to any individual density bin. Furthermore, the range of
radii contributing to a density bin becomes even broader near the
center of the halo due to the shallower profile of the inner
region. The core of the $N_{200}=10^{6}$ halos are resolved better
relative to the halos sampled with $N_{200}=10^{2}$. The halos with
$N_{200}=10^{2}$ particles tend to underestimate the densities near
the center. Here the boost due to scatter in densities is somewhat
compensated by the poor resolution, enabling it to nicely but arguably
misleadingly recover the analytic mass profile even near the center of
the halo.

\subsection{The one-point distribution function in the Millennium and
  Millennium-II simulations}  \label{subsec3}

In the halo model, it is assumed that all the matter in the Universe
resides in a halo of some mass. With this assumption, the whole matter
distribution can be represented as a superposition of a set of halos
in different mass ranges. To define the model one only needs to
specify the density profiles of halos and the halo mass function. The
density profile of dark matter halos can be described by the NFW
profile (equation \ref{eq:nfw}), which is a function of radius and
mass of the halo. The concentration $c={r_{200}}/{r_{s}}$ depends
weakly on halo mass, and assuming the mass-concentration relation is
known, one can write down the density $\rho(r, M)$ for any particle
residing at a radius $r$ of a halo of mass $M$. For a smooth NFW halo
the probability distribution function $P(\rho)$ is simply given by the
fraction of the volume at density $\rho$, i.e.
\begin{equation}
P(\rho)= \frac{1}{V} \frac{{\rm d}V}{{\rm d}\rho}=
\frac{3}{r_{200}^3} \, r^2 \, \frac{{\rm d}r}{{\rm d}\rho}.
\end{equation}
For an NFW halo,
$\frac{d \ln \rho}{d \ln r} = \frac{r_{s}+3r}{r_{s}+r}$. So for a given
mass of the halo and a specified mass concentration relation one can
analytically calculate the density probability distribution function of
the halo.

\begin{table}{}
\begin{center}
\begin{tabular}{|c|c|c|}
\hline
Halo mass range&  Fraction ($\%$) & Fraction ($\%$)\\
( in $\, h^{-1} \, M_{\odot}$)&  in MS & in MS-II\\
\hline
$10^{9}-10^{10}$  &$0$& $1.721$ \\
$10^{10}-10^{11}$ &$0$& $6.94$ \\ 
$10^{11}-10^{12}$ &$2.182$& $7.90$ \\ 
$10^{12}-10^{13}$ &$4.495$& $8.391$ \\
$10^{13}-10^{14}$ &$5.981$& $9.103$ \\
$10^{14}-10^{15}$ &$7.639$& $10.435$ \\ 
$10^{15}-10^{16}$ &$9.972$& $$ \\ 
\hline
\end{tabular}
\end{center}
\caption{Fraction of particles in each halo mass bin that 
  account for the excess in the actually measured distribution
  function in the corresponding halo mass bin relative to the
  theoretical distribution functions.}
\end{table}

\begin{figure*}
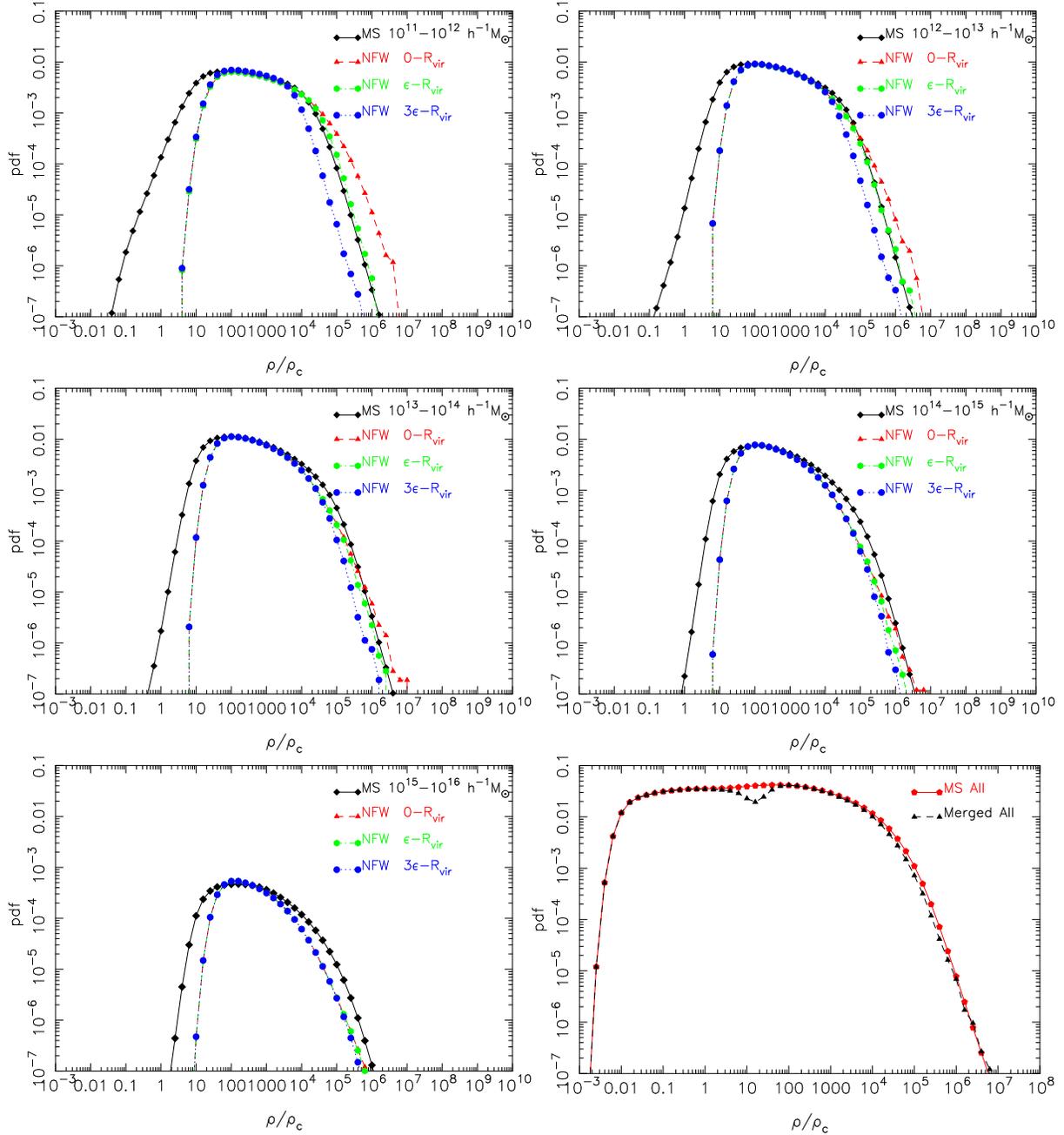

\resizebox{8cm}{!}{\rotatebox{-90}{\includegraphics{mill_11t12.ps}}}%
\resizebox{8cm}{!}{\rotatebox{-90}{\includegraphics{mill_12t13.ps}}}\\
\resizebox{8cm}{!}{\rotatebox{-90}{\includegraphics{mill_13t14.ps}}}%
\resizebox{8cm}{!}{\rotatebox{-90}{\includegraphics{mill_14t15.ps}}}\\
\resizebox{8cm}{!}{\rotatebox{-90}{\includegraphics{mill_15t16.ps}}}%
\resizebox{8cm}{!}{\rotatebox{-90}{\includegraphics{mill_combined.ps}}}\\
\caption{The top left panel shows the one-point density distribution
  function (black squares) of MS particles residing in halos in the
  mass range $10^{11}-10^{12}\,h^{-1} {\rm M}_{\sun}$. For comparison,
  the one-point distribution computed from Monte Carlo realizations of
  a NFW halo with mass $10^{12}\,h^{-1} M_{\sun}$ is scaled according
  to the total number of particles present in halos in the mass range
  $10^{11}- 10^{12}\,h^{-1} M_{\sun}$. The distributions computed by
  restricting particles to within $r\in [0, r_{\rm vir}]$,
  $[1\epsilon, r_{\rm vir}]$ and $[3\epsilon,r_{\rm vir}]$ are shown
  in different colors and symbols. The other panels, except for the
  one in the bottom right, show the results for halo mass ranges
  $10^{12}-10^{13}$, $10^{13}-10^{14}$, $10^{14}-10^{15}$ and
  $10^{15}-10^{16}\,h^{-1} M_{\sun}$, respectively. Finally, the
  bottom right panel shows the sum of scaled distributions of NFW
  halos in different mass bins, and adds it to the distribution of all
  the non-halo particles in the simulation.  The full one-point
  distribution of all the particles in the MS is also plotted (black
  triangles), for comparison. We note that while summing the
  distributions of Monte Carlo NFW halos over different halo mass bins
  we considered only particles within $r\in [1\epsilon,r_{\rm vir}]$
  of the halo for the two lowest mass bins. For the rest of the bins
  all particles are considered.}
  \label{fig:5}
\end{figure*}

\begin{figure*}
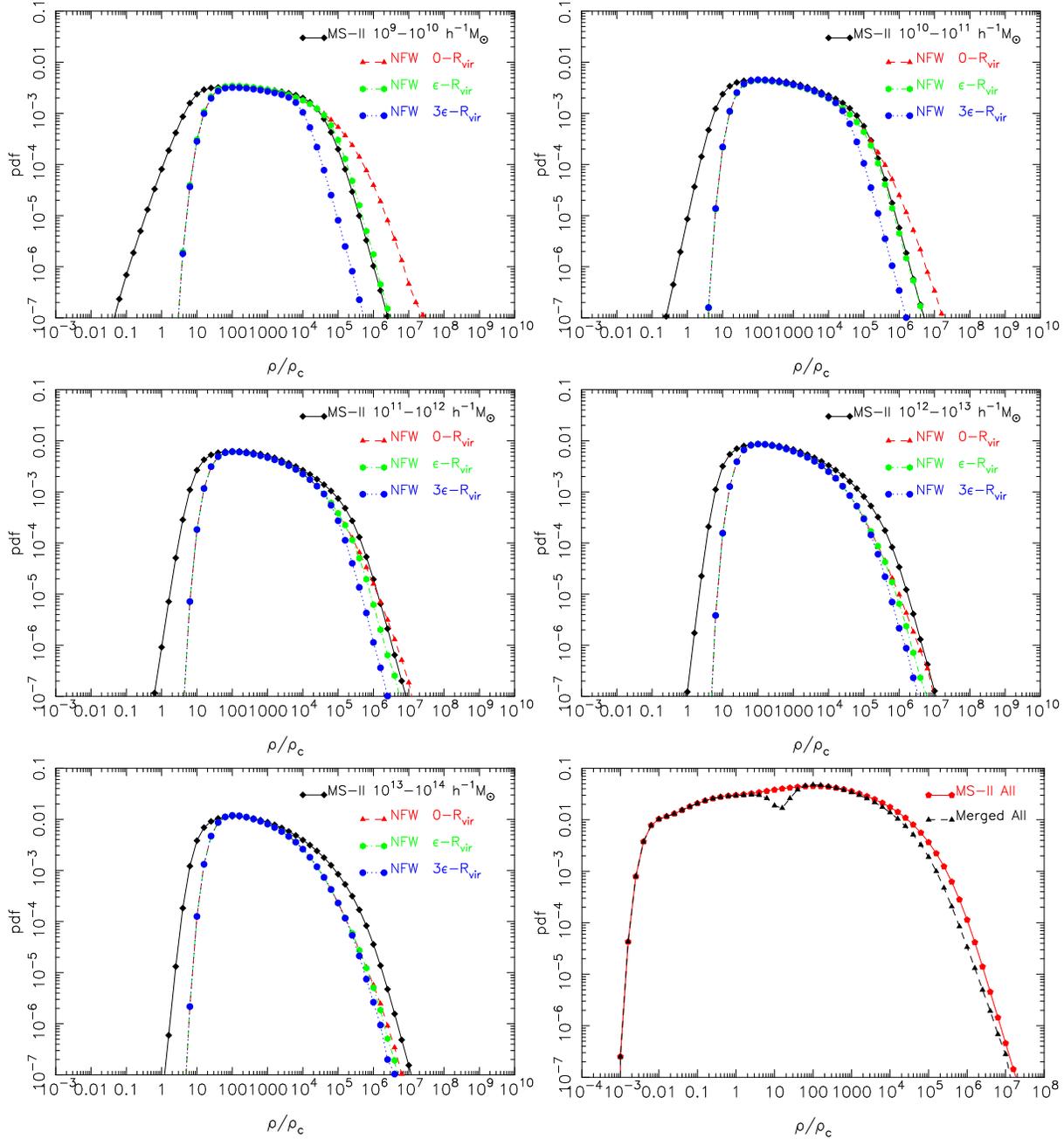

\resizebox{8cm}{!}{\rotatebox{-90}{\includegraphics{mill2_9t10.ps}}}%
\resizebox{8cm}{!}{\rotatebox{-90}{\includegraphics{mill2_10t11.ps}}}\\
\resizebox{8cm}{!}{\rotatebox{-90}{\includegraphics{mill2_11t12.ps}}}%
\resizebox{8cm}{!}{\rotatebox{-90}{\includegraphics{mill2_12t13.ps}}}\\
\resizebox{8cm}{!}{\rotatebox{-90}{\includegraphics{mill2_13t14.ps}}}%
\resizebox{8cm}{!}{\rotatebox{-90}{\includegraphics{mill2_combined.ps}}}\\
\caption{The same as Figure~\ref{fig:5}, but for the MS-II.}
  \label{fig:6}
\end{figure*}

We now contrast this expected density distribution with actual
measurements for the MS and MS-II when the density field is
constructed with the DTFE.  The probability distribution functions for
all the dark matter particles at redshift $z=0$ from the MS and MS-II
are shown as black curves in the two panels of Figure \ref{fig:4}. The
plots show that with the DTFE we are able to recover density values
spanning about $\sim 11$ orders of magnitude, from underdense voids to
the fully collapsed halos. The distribution is flat over nearly six
orders of magnitude in density, and it does not exhibit any apparent
linear to non-linear transition. The lower limits of the distribution
functions shown in these figures have a Poisson error of $\sim 3\%$.

To isolate the contribution from collapsed halos we consider dark
matter particles according to the masses of their host FoF halos. We
identify the host halos by cross-matching particles IDs with FoF group
IDs in the simulation. We also identify all the particles which are
not part of any FoF halo with $20$ or more particles. The density
distributions for each of these cases are computed separately. The
results for the different components are shown together in
Figure~\ref{fig:4}. It should be noted that even though the low mass
halos are more concentrated compared to their high mass counterparts,
the distribution systematically extends to higher density values for
the higher mass halos. This is opposite to what one would naively
expect. This result could be explained by the fact that the smaller
mass halos are sampled with fewer particles which poorly resolve the
highly concentrated cores in these halos.  In addition, the effect of
gravitational softening (which introduces a soft core in the halo) in
N-body simulations is expected to be more severe for less massive
halos as the softening length is a relatively larger fraction of their
virial radii. The results for the MS-II in the right panel are quite
similar, except for the fact that the overall shape of the
distribution function at intermediate densities is somewhat different
than that found for the MS. The distribution in this case is curvier
and not quite as flat. This presumably reflects the larger fraction of
particles bound in halos in the MS-II compared with the MS. It is also
noticeable that at a given halo mass the density distributions extend
to high densities in MS-II. Again this generally reflects the larger
softening of the MS.

More detailed comparisons between MS and MS-II are shown in
Figure~\ref{fig:7}. As the DTFE does not use any specific length scale
for smoothing it is assured that the truncation in the tails of the
distribution is not a result of a spatial resolution limit imposed by
the density estimator. However, the intrinsically limited volume and
mass resolution of the simulation is expected to introduce an
undersampling of the tails of the distribution \citep{colombi94,
  bagla05}. Smaller volumes undersample rare events in both overdense
and underdense regions. A similar effect is caused by lower mass
resolution primarily due to its limitation in resolving lower mass
halos and smaller voids. MS-II has $125$ times better mass resolution
and $5$ times smaller softening length than MS enabling it to resolve
the highly concentrated smaller mass subhalos. It can be seen in the
top left panel of Figure~\ref{fig:7} that the distributions in MS and
MS-II are quite similar, apart from the fact that the distribution for
the MS-II extends slightly more on both the low density and high
density ends. The effects of finite volume and finite mass resolutions
somewhat compensate each other in the MS and MS-II but the slightly
extended tail of the distribution at the both end in the MS-II suggests
that the combined effects of finite mass resolution and gravitational
softening are more important than the effect of finite volume. 

The results for the non-halo particles are shown in the top right
panel. The distribution in the MS-II has lower amplitude than the MS
in intermediate density ranges because of its higher mass resolution
and smaller softening, which allows it to resolve more bound objects
at small mass. It is also interesting to note that the high density
tail of the distribution for the non-halo particles extends up to
$\sim 10^{6} \, \rho_{c}$ suggesting the presence of some high density
sites even outside the bound FoF halos. The distributions in different
individual halo mass bins in MS and MS-II are shown in the two middle
and two bottom panels. The MS-II distributions have a sharper
low-density cut-off and a more extended high density tail than in the
MS in all cases. Any halo of a particular mass is sampled with $125$
times more particles in MS-II than in MS. This gives MS-II better
power to resolve small subhalos and the boundary of the FoF
groups. Further, the softening length in MS-II is $5$ times smaller
than in the MS reducing softening effects.  These effects are more
dominant in low mass halos and can be clearly seen as larger shifts in
the lower halo mass bins. We note that we use dark matter halos
    identified with the FoF algorithm \citep{davis}. There are also
    other halo finding algorithms, for example based on the spherical
    overdensity (SO) approach \citep{press}, the adaptive grouping of
    particles around density peaks \citep{eisenstein98, neyrinck05},
    or the phase-space distribution of dark matter particles
    \citep{diemand06, maciejewski09, falck10}. The halo boundaries in
    general depend on the free parameters in the corresponding
    algorithms (e.g. linking length in FoF, density cut-off in SO),
    and the individual halos identified with different methods can
    sometimes vary substantially. But fortunately the mean properties
    of the dark matter halos agree quite well regardless of the chosen
    algorithm, and the differences in the halo mass functions are also
    quite small \citep{jenkins01, knebe11}.

In the two top panels of Figure~\ref{fig:8}, we compare our analytical
halo model for the one-point distributions of the density summed over
different halo mass bins against that directly obtained for the MS and
MS-II. For MS and MS-II, we sum the results for $5$ and $6$ equally
spaced different halo mass bins each spanning a decade in the halo
mass ranges $10^{11}- 10^{16}\,h^{-1} M_{\sun}$ and $10^{9}-
10^{15}\,h^{-1} M_{\sun}$, respectively. We also show the ratio of the
one-point distributions from the N-body data and the analytical model
in the left panel of Figure~\ref{fig:8a}. The excess in the observed
distribution compared to the model prediction is higher in the MS-II
than in the MS which is most likely related to the relative abundance
of substructures. We explore this issue in detail in the remaining
part of the paper. One can also clearly see a larger suppression of
the high density tail of the distribution in the MS compared to the
MS-II due to its larger softening length. The sharp drop in the
analytical predictions in the low density regime corresponds to a
truncation of the halos at their virial radii. In the bottom two
panels of Figure~\ref{fig:8}, we also compare the analytical
predictions for the one-point distribution function of NFW halos
summed across the different halo mass bins used in the analysis, and
the combined one-point distribution function of DTFE densities
computed from Monte Carlo NFW halos across the same halo mass bins.
The ratio of the one-point distributions from the Monte Carlo analysis
and the analytic model are shown in the right panel of
Figure~\ref{fig:8a}. We use the same mass-concentration relation as
employed for the analytic estimates.  It should be noted that we are
not using any fit for the halo mass function in our analytical
model. The one-point distributions for the Monte Carlo NFW halos with
different masses are scaled according to the total number of particles
present in different halo mass bins, as directly measured in the
simulations, and then summed up over all the halo mass bins. Our
analytical model is based on equation~(7) combined with the fact that
the results for each halo mass bin are scaled by exactly the same
amount as their Monte Carlo counterparts. The results for different
halo mass bins are then summed up. We see that the analytical
one-point distribution function is quite well described by the results
from Monte Carlo simulations, apart from the fact that the analytical
predictions show a sharp drop in the distribution function due to the
truncation of all NFW halos at $r_{200}$. The DTFE densities do not
show this sharp drop due to the Poisson sampling of the halos. The
slightly more extended high density tail seen in the MS-II comes from
its ability to incorporate lower mass halos which are more
concentrated. Thus, the DTFE traces the analytical one-point
distribution function of the densities quite well, and the amount of
excess (substructure) and the shape of the high density tail of the
distribution in the N-body simulations are governed by the mass
resolution, gravitational softening and simulated volume.

We have also modeled the one-point distributions for different mass
bins in the N-body simulations with ideal spherical NFW halos of similar
masses. For this we generate $10^4$ , $10^3$, $10^2$ , $10$ and $1$
Monte Carlo realizations of NFW halos with masses $10^{12} \, h^{-1}
\, M_{\odot} $, $10^{13} \, h^{-1} \, M_{\odot}$, $10^{14} \, h^{-1}
\, M_{\odot}$, $10^{15} \, h^{-1} \, M_{\odot}$ and $10^{16} \, h^{-1}
\, M_{\odot}$, respectively. We use the mass-concentration relation
given by \citet{neto} and the particle mass $8.61 \times 10^8 \,
h^{-1} \, M_{\odot}$ of the MS. The concentration of halos depends
very weakly on mass, and for simplicity we assume that the
concentration does not change significantly within each of the narrow
halo mass ranges. Different numbers of halos for the different mass
ranges are constructed to account for the effects of discreteness
i.e.~to incorporate the fact that a $10^{n_{1}} \, h^{-1} \,
M_{\odot}$ halo is resolved with $10^{{n_{1}}/{n_{2}}}$ times more
particles than a $10^{n_{2}} \, h^{-1} \, M_{\odot}$ halo for any
given values of $n_{1}$ and $n_{2}$ (where $n_{1} > n_{2}$). We use
all the density estimates from $10^{{n_{1}}/{n_{2}}}$ simulated NFW
halos with mass $10^{n_{2}} \, h^{-1} \, M_{\odot} $ to compute their
one-point distribution, ensuring that the same number of density
estimates are used to determine the one-point distribution of density
for NFW halos with different masses.  We followed the same approach
for the MS-II as well, except that here we can go down another
two decades in halo mass. We use the same mass-concentration relation
for the MS-II, extrapolated to lower halo masses as needed.

We compute in this way the one-point distribution of Monte Carlo NFW
halos with different masses, and then scale them according to the
total number of particles present in different halo mass bins as
directly measured in the MS, in order to compare the theoretical
predictions of this simple halo model with the measurements directly
obtained from the MS (Figure~\ref{fig:5}). It should be noted that we
have not used the halo mass function to model the one-point
distribution here. Instead, we have used the total number of particles
in a halo mass bin to predict the expected one-point distribution from
halos in that bin. The results of this comparison for the MS and the
MS-II are shown in Figures~\ref{fig:5} and \ref{fig:6}, respectively.

We find that the one-point distribution in each mass bin is reproduced
nicely at intermediate densities when the results for the Monte Carlo
NFW halos in that mass bin are scaled according to the total number of
particles found in the N-body simulation in the same mass bin
(Figures~\ref{fig:5} and \ref{fig:6}). The low density part of the
distribution in each mass bin is however missed by the theoretical toy
model. This is to be expected as we are only considering the particles
within the virial radii of the spherical mock NFW halos, whereas in
reality the FoF halos in N-body simulations extend beyond their virial
radii and generally have quite irregular shapes near their edges. At
higher densities, the lower mass bins show higher values of the PDF
for the theoretical one-point distribution compared with the N-body
simulations. This is because the mock NFW halos do not involve any
softening whereas a gravitational softening is present in the N-body
simulations from where the FoF halos are identified. In order to
explicitly check for the impact of the softening length ($\epsilon$)
in N-body simulations we have made an experiment where we prevented
that particles in the mock halos are placed within $1 \epsilon$ from
their centers when computing the one-point distribution for all NFW
halos in each mass bin. Interestingly, when we limit the particles to
the radii outside of the softening range, the high density tail is
nicely consistent between the N-body simulations and the mock
halos. It thus seems clear that softening primarily affects the high
density tail of the one-point distribution by suppressing the highest
density values. This simple explanation does not work as well for the
highest halo mass bins, presumably because their structure is less
strongly affected by the softening and they feature much more halo
substructure.

It is interesting to note that the one-point distribution in the
higher halo mass bins shows an excess over the theoretical
prediction. The fractions of particles in each halo mass bin which
account for this excess are different and increase with increasing
halo masses. Specific numbers for our measurements are reported in
Table~2. In a smooth NFW halo, there are no substructures whereas the
halos formed in N-body simulations host numerous small subhalos that
typically account for a few percent up to $\sim 15\%$ of the mass. The
excess we find in the one-point distributions is most likely the
direct consequence of the presence of substructures in the massive
dark matter halos formed in N-body simulations. The excess accounts
for up to $\sim 8-10\%$ of the total particles in the highest halo mass
bin, consistent with typical substructure mass fractions. The excess
decreases with halo mass and is completely absent or almost negligible
in the lowest mass bins. Again, the decrease in the amount of
substructures with decreasing halo mass is consistent with previous
findings \citep{gao} and with expected numerical resolution
limitations. Further, it is to be noted that the MS-II shows a higher
substructure abundance compared to the MS in each halo mass bin. This
is related both to the higher mass resolution and to the smaller
softening length in the MS II which enables it to resolve lower mass
subhalos.

\begin{figure}
\resizebox{8cm}{!}{\rotatebox{-90}{\includegraphics{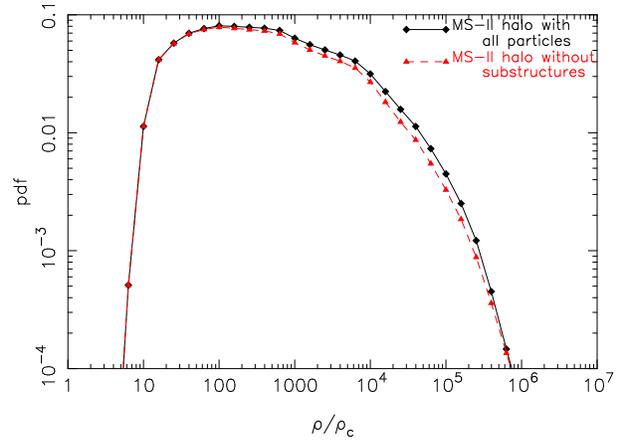}}}%
\caption{The one point distribution of density for a halo from the
  MS-II for all the particles residing within $1.5 \, r_{200}$ around
  the halo centre. For comparison, the red dashed line shows the
  one-point distribution of the same halo after all its substructures
  within $1.5 \, r_{200}$ has been removed. The virial radii of this
  halo is $0.60 \, h^{-1} \, {\rm Mpc}$, and a total of $8079161$
  particles reside within $1.5 \, r_{200}$, out of which $528753$
  ($\sim 6.5 \%$) are subhalo members.}
  \label{fig:6a}
\end{figure}

The effect of substructures on the one-point distribution is shown
explicitly in Figure~\ref{fig:6a}, which focuses on a well resolved
halo from the MS-II. The substructures in the MS-II halo are here
identified with the {\small SUBFIND} algorithm \citep{springel01} and
are available from the MS-II database. We identified all the particles
within $1.5 \, r_{\rm vir}$ around the center of that halo and computed
the one-point distribution of the density. We then removed all the
substructure particles within that radius and computed the
distribution function again. The results are compared in
Figure~\ref{fig:6a}, which clearly highlights the excess due to the
presence of substructures in the halo. The substructures constitute
$\sim 6.5 \%$ of the total particles in this halo within our chosen
radius, and this fraction is consistent with the values listed in
Table~2.

\begin{figure}
\resizebox{8cm}{!}{\rotatebox{-90}{\includegraphics{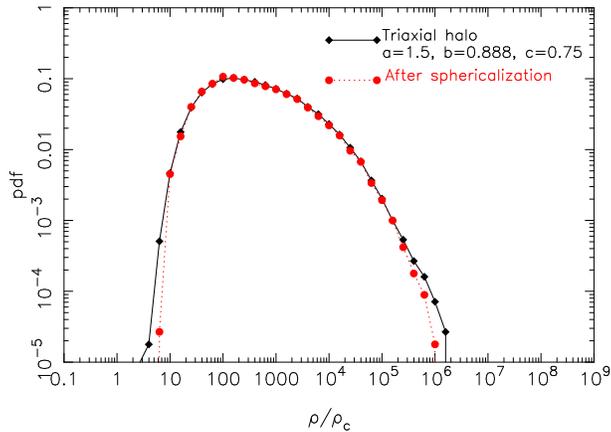}}}
\caption{One-point density distribution function of a Monte Carlo
  realization of a triaxial NFW halo (black) of mass
  $M_{200}=10^{14}\,h^{-1} M_{\sun}$, represented with particle mass
  $8.61 \times 10^{8}\,h^{-1} M_{\sun}$. The density distribution
  function of the same halo after sphericalization is shown in red. To
  avoid any boundary effects only the particles inside $r_{200}$ are
  used in both cases.}
  \label{fig:9}
\end{figure}

One should also keep in mind that the dark matter halos in simulations
are not exactly spherical. Rather their shapes resemble in general
triaxial ellipsoids with a preference for prolate configurations
\citep{dubin, cole, jing, vogelsberger}. The shapes of the majority of
the halos forming in N-body simulations are characterized by a mean
axis ratio of about $1:0.74:0.64$ \citep{kasun, bailin, allgood}.
Fitting a NFW profile to such halos inevitably involves spherical
averaging. The spherical averaging of a triaxial halo will introduce
systematic differences in the density estimates in different parts of
the halo. Some parts of the halo have in reality densities which are
larger/smaller than the spherically averaged densities we try to
reproduce in our mock models. This deviation in the densities could be
quite high depending on the triaxiality of the dark matter halos and
it is important to investigate if the excess in the observed one-point
distribution of density is related to this issue. In order to test
this we simulated a triaxial NFW halo of mass $M_{200} = 10^{14} \, h^{-1}
\, M_{\odot}$ with axis ratios $a:b:c=1.5:0.888:0.75$. To mimic the
effect of sphericalization we randomize the azimuthal co-ordinates
($\theta$, $\phi$) of all the particles in the halo. The one-point
distribution function of the triaxial halo before and after
sphericalization are compared in Figure~\ref{fig:9} where we can see
that this effect does not make a significant change in the one-point
distribution function of density despite our choice of a deliberately
extreme (yet still possible) axis ratio. This result suggests that the
observed excesses in the one-point distributions (Figures \ref{fig:5}
and \ref{fig:6}) are a direct outcome of the substructures present in
the dark matter halos, and that the halo shape plays only a very
subdominant role.

As an aside, we note that the dependence of substructure abundance on
mass resolution has important implications for predictions of the
expected extragalactic gamma ray background hypothetically caused by
the self-annihilation of dark matter particles in dark matter halos.
The rate of WIMP annihilation and hence the intensity of the
annihilation radiation in a dark matter halo is proportional to the
volume integral of the square of the dark-matter density. The
existence of lumps or subhalos in dark matter halos is expected to
enhance the annihilation rate by a significant boost factor
\citep{bergstrom, stoehr, koushi, pieri, kuhlen08, springel08,
  kuhlen09, kamion10}. The smallest subhalos are the densest sites in
the halo and the predicted boost factor depends significantly on how
well these subhalos are resolved in the N-body simulations. Given that
halos in the MS-II host more resolved substructures than the MS
indicates that the theoretical prediction of the intensity of
annihilation signal from any simulated dark matter halo depends on the
resolving power imposed by the softening length and by the finite mass
resolution. One has to correct for this by extrapolating the relations
between substructure abundance and mass resolution (at a fixed
softening length) to lower particle masses in order to make a
prediction about the signal expected in reality.

We combine the results of our Monte Carlo simulations of NFW halos in
different mass bins, and add also the one-point distribution of the
non-halo particles found in simulations to obtain a halo model for the
full one-point distributions in the MS and MS-II. The results are
shown in the bottom right panel of Figures~\ref{fig:5} and
\ref{fig:6}. The dip in the middle of the distributions results from
the truncated spherical boundaries of the idealized NFW halos, which
does not take into account the fact that the real halos in simulations
are far more irregular at their boundaries and extend to lower
densities as well. At higher densities, there are nevertheless also
differences between the direct results of the N-body simulations and
the Monte Carlo model. Clearly, larger differences are seen in the
case of the MS-II compared with the MS, an effect that we attribute to
the higher abundance of substructures in the MS-II.

\begin{figure}
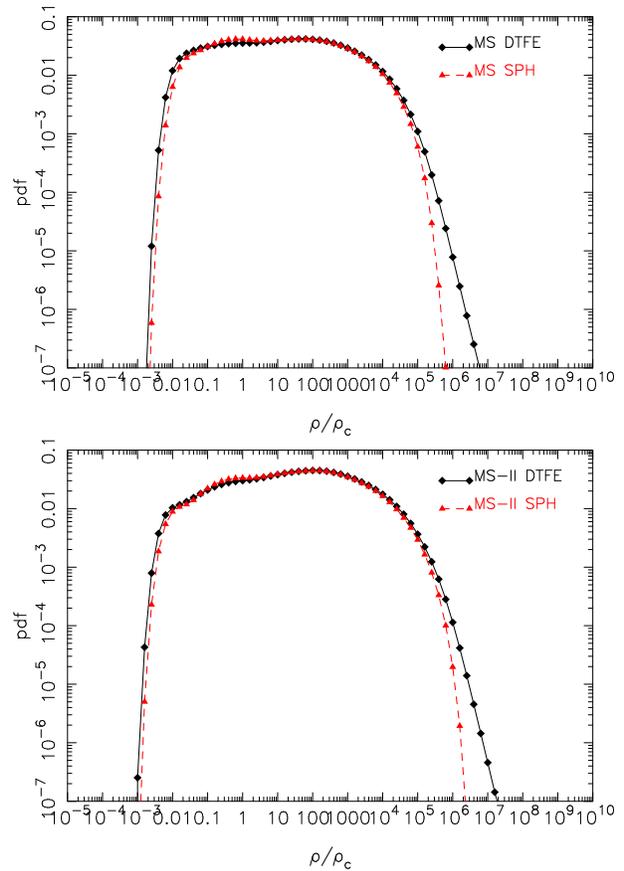

\resizebox{8cm}{!}{\rotatebox{-90}{\includegraphics{msdtsph.ps}}}\\
\resizebox{8cm}{!}{\rotatebox{-90}{\includegraphics{ms2dtsph.ps}}}
\caption{Comparison of the density distributions measured with DTFE
  and SPH smoothing. The top and bottom panels show results for the MS
  and the MS-II, respectively.}
  \label{fig:7a}
\end{figure}

\begin{figure} 
\resizebox{8.5cm}{!}{\rotatebox{0}{\includegraphics{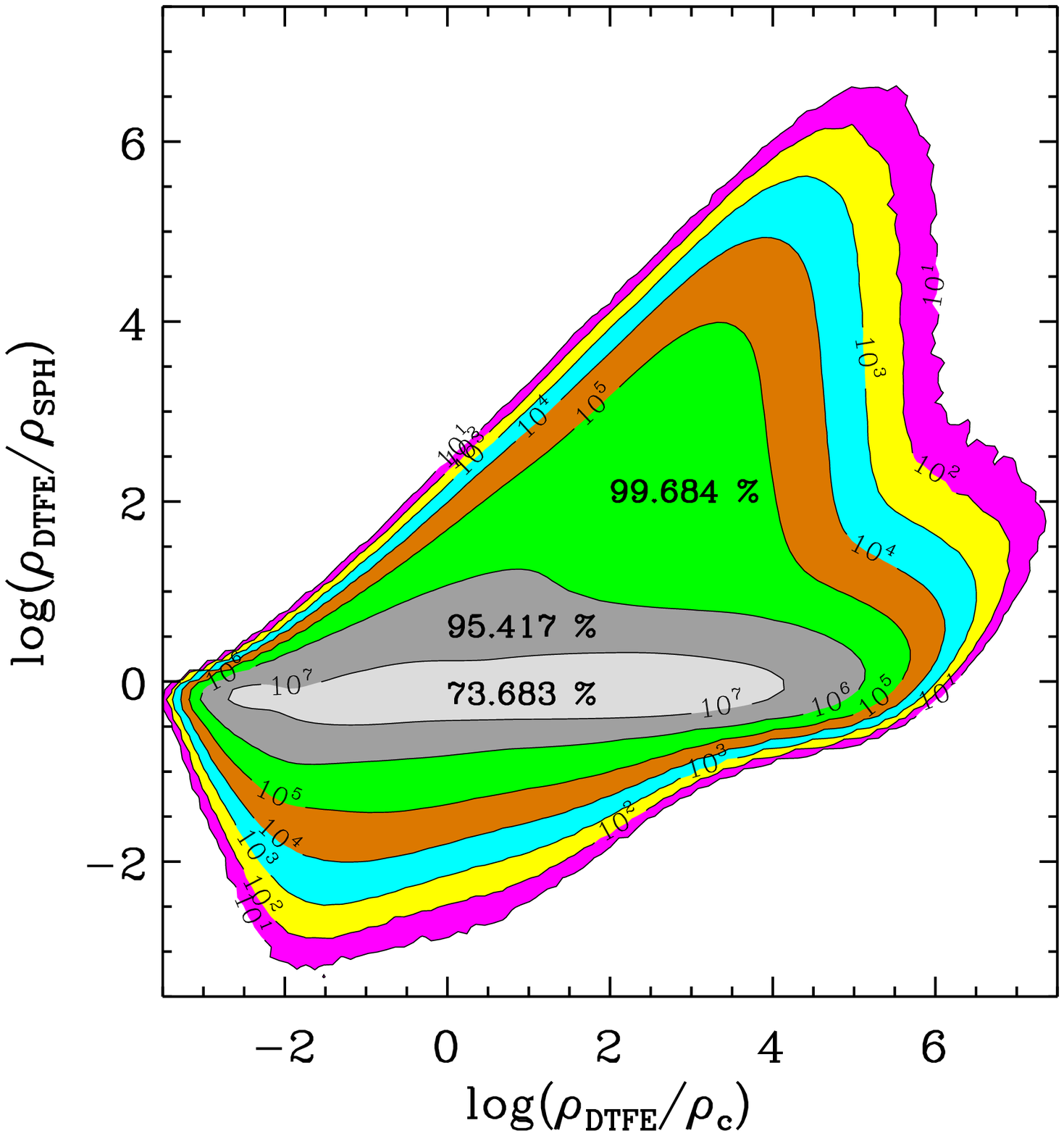}}} 
\caption{Two-dimensional histogram of the ratio of DTFE to SPH density
  estimated for the MS-II, as a function of the DTFE density.  A total
  of $100 \times 100$ pixels are used to estimate the
  histogram. Different contours correspond to the pixels with a count
  equal to the numbers used for labeling the respective
  contours. Different coloured regions mark the pixels with counts
  intermediate to their bounding contours. The numbers shown inside
  the coloured regions give the total cumulative counts in all the
  pixels contained inside the outer contours of the corresponding
  regions. This plot clearly shows the systematic overestimate and
  underestimate of densities by SPH over DTFE at low and high
  densities, respectively.}
  \label{fig:7ratio}
\end{figure}

Finally, we compare the distribution of DTFE and SPH densities from
the MS and MS-II in Figure~\ref{fig:7a} to check how the choice of
density estimator influences the tails of the distribution. The
smoothing lengths in SPH are chosen such that the sum involves $32$
nearest neighbours. This figure shows that for the lowest density
regions or voids the DTFE and SPH smoothing give very similar
results. At very low densities, the slightly higher values found in
the DTFE distribution come probably from a more accurate
representation of void boundaries than in SPH. Similarly, the extended
kernel of SPH smoothes out high density regions like filaments or halo
centers leading to an underestimate of the PDF in such regions. In
both the MS and MS-II there is a small bump in the density range
$0.1-5 \rho_{c}$ displaying an excess of the SPH density estimate as
compared to the DTFE density measurement. 

In Figure~\ref{fig:7ratio}, we show how the DTFE and SPH density
estimates compare with each other at different densities in the
MS-II. It can be clearly seen that for most of the particles, DTFE and
SPH give comparable results, while residual systematic differences
show up as oppositely skewed tails at high and low densities. These
systematic differences in low and high density regions account for the
differences in the SPH and DTFE density distributions.  A detail
comparison between DTFE and SPH is given in \citet{pelupessy} and our
findings are consistent with theirs. Recently, \citet{abel} pointed
out that even the VTFE underestimates the densities in regions around
filaments and sheets, as compared to a novel technique that more
accurately represents dark phase-space sheets. So besides factors like
mass resolution, gravitational softening and simulated volume, the
choice of density estimator plays a crucial role in shaping the tails
of the density distribution.

\section{Conclusions} \label{secconclusions}

The present analysis shows that the part of the one-point distribution
function represented by collapsed halos can be quite well described by
a simple superposition of a set of NFW halos over different mass
ranges. However, the one-point distribution functions in N-body
simulations also show a prominent hump when individual halo mass bins
are considered, especially for the more massive halos. This excess
with respect to the distribution obtained for smooth NFW halos
originates in the substructures present in the massive dark matter
halos. The amount of resolved substructures depends on the mass of the
halo, and especially on the finite mass resolution of the N-body
simulation. Further, the gravitational softening suppresses the high
density tail of the one-point distribution in halos, introducing a
soft core which is more noticeable in smaller mass halos. Both of
these effects imply that the high-density tail is still underestimated
both in the direct N-body simulations and the analytical halo model.

We find that finite simulation volume, finite mass resolution,
gravitational softening as well as the method for estimating the
density field all influence the tails of the measured one-point
density distribution. We note that this distribution function is a
particularly important simulation prediction, as it, for example,
determines the intensity of the WIMP annihilation signal from a
representative volume, which sensitively depends on the ability to
resolve the abundant yet dense small-mass structures. Our analysis
with the DTFE in the MS and MS-II suggests that the effect of finite
mass resolution and gravitational softening are the primary
limitations rather than a finite simulation volume. Also, it is
worthwhile to employ the DTFE techniques instead of simpler schemes
for density reconstruction such as SPH-like smoothing, due to its
sharper resolving power.

 
\section*{Acknowledgments}

The Millennium simulations used in this study where carried out at the
Max-Planck Computing Centre (RZG) in Garching, Germany. BP thanks the
Alexander von Humboldt Foundation for financial support. VS
acknowledges partial support through TR33 `The Dark Universe' of the
German Science Foundation (DFG).

 \bibliographystyle{mn2e}

\end{document}